\documentclass{optica-article}
\journal{opticajournal}
\articletype{Research Article}
\usepackage{lineno}
\usepackage{mathtools}
\usepackage{orcidlink}

\begin{document}

\title{Automating Experimental Optics with Sample Efficient Machine Learning Methods}

\author{
Arindam Saha\authormark{1, *} \orcidlink{0000-0002-3120-3375},
Baramee Charoensombutamon\authormark{2} \orcidlink{0000-0002-8317-0632},
Thibault Michel \authormark{1} ,
V. Vijendran \authormark{1, 3} \orcidlink{0000-0003-3398-1821},
Lachlan Walker \authormark{4} ,
Akira Furusawa \authormark{2} \orcidlink{0000-0002-6176-6742},
Syed M. Assad \authormark{1, 3} \orcidlink{0000-0002-5416-7098},
Ben C. Buchler \authormark{1} \orcidlink{0000-0002-2852-7483},
Ping Koy Lam \authormark{1, 3} \orcidlink{0000-0002-4421-601X},
and Aaron D. Tranter \authormark{1, *} \orcidlink{0000-0002-0582-8275}
}

\address{
\authormark{1}Department of Quantum Science and Technology, Australian National University, 38a Science Rd, Acton, 2601, ACT, Canberra, Australia\\
\authormark{2}Department of Physics, University of Tokyo, Bunko City, 113-8654, Tokyo, Japan\\
\authormark{3}A*STAR Quantum Innovation Centre (Q.InC),
Agency for Science, Technology and Research, 2 Fusionopolis Way, 
08-03 Innovis 138634, Singapore\\
\authormark{4}2pi Software, 903/50 Clarence Street, Sydney, 2000, NSW, Australia
}

\email{\authormark{*}arindam.saha@anu.edu.au, aaron.tranter@anu.edu.au}

\begin{abstract*}
    As free-space optical systems grow in scale and complexity, troubleshooting becomes increasingly time-consuming and, in the case of remote installations, perhaps impractical. An example of a task that is often laborious is the alignment of a high-finesse optical resonator, which is highly sensitive to the mode of the input beam. In this work, we demonstrate how machine learning can be used to achieve autonomous mode-matching of a free-space optical resonator with minimal supervision. Our approach leverages sample-efficient algorithms to reduce data requirements while maintaining a simple architecture for easy deployment. The reinforcement learning scheme that we have developed shows that automation is feasible even in systems prone to drift in experimental parameters, as may well be the case in real-world applications.
\end{abstract*}

\section{Introduction}

Optical systems are the backbone of modern-day information transmission, providing global connectivity and data transmission rates unmatched by competing technologies. Free-space optics (FSO) allows long-haul transmission of information over channels that are far less congested than traditional radio-frequency (RF) bands and removes the need for costly optical fibre infrastructure \cite{trichili_roadmap_2020, malik_free_2015}. Notably, FSO have also been shown to outperform RF communications in near-Earth space-based settings \cite{toyoshima2007comparison} with demonstrated high-speed link capability \cite{walsh_demonstration_2022, kaur2024400}. Outside of established communication technology, FSO have an important role to play in implementations of optical quantum computing \cite{OBrien2009, Humphreys2013, Budinger2024}, quantum communications \cite{liao_satellite--ground_2017,takenaka_satellite--ground_2017,gunthner_quantum-limited_2017, Jaouni2025}, quantum sensing \cite{Fang2024, Aslam2023} and power-efficient information processing \cite{hu_diffractive_2024, wang_optical_2022, zhou_large-scale_2021}. A persistent challenge in FSO systems is the precise alignment required to maintain high operating performance. As the complexity of FSO schemes increase, maintaining alignment becomes more burdensome for human operators and in certain circumstances may not even be possible for remotely deployed systems, including space. This has justifiably sparked an interest in autonomous control \cite{acernese_virgo_2006, romann_automatic_2005, Brown2021, Fulda2017, Morrison1994, tarquin_ralph_automatic_2020, Rakhmatulin2024, Hinrichs2020, Mukund2023, Mareev2023, Sun2020, interferobot, Praeger2021, lea-vicky2024, Qin2025, Vazquez2021}.

The widespread utility of machine learning (ML) methods in optimisation and control stems from their ability to learn relationships in observable data. This sidesteps the need to develop handcrafted feature extraction, a task that is often difficult or infeasible for complicated physical systems.

ML methods often rely on large amounts of training data to learn complex relationships. This sort of methodology can be impractical in experimental contexts where data acquisition is costly and time-consuming. For applications in physical systems, it is highly advantageous to use sample-efficient methods, such as ensemble learning \cite{Sagi2018}, model-based Reinforcement Learning (RL) \cite{MBRLSurvey}, and transfer learning \cite{Pan2010}.

While ML techniques are now ubiquitous in control and optimisation applications, there has been comparatively little work on their use in precision optical alignment \cite{Rakhmatulin2024}. Automation of precision alignment has often involved precise calibration of controls and other system parameters. This can be a limiting factor if the system parameters are prone to drift, often necessitating more expensive precision actuators \cite{Rakhmatulin2024,sanchez2018control}. If the system exhibits minimal or slow drift, online-learning using a model-free approach is often effective and has been demonstrated in a number of optical setups, including optical fibre alignment \cite{murakawa_automatic_2004, mathew2021raspberry} and optical cavities \cite{Qin2025}.

Notably, \cite{Qin2025} achieved a high coupling efficiency and alignment speed that exceeds experienced experimentalists, employing a mode identification system based on convolutional neural networks and online parameter optimisation with a genetic algorithm. Generally, however, genetic algorithms do not scale well with complexity and dimensionality and require careful tuning. 
Deep learning (DL) techniques offer a more flexible approach to online-learning at the expense of higher computational cost while still side-stepping algorithmic tuning \cite{Tranter2018}. DL techniques have been used to control various optical systems, for example, wavefront correction \cite{Hinrichs2020}, X-ray generation \cite{Mareev2023} and optical tweezers \cite{Praeger2021}.

Any method that assumes a static mapping between a set of control parameters and the response of the system will quickly become ineffective if the experimental parameters drift. For optical systems, hysteresis in the actuators and thermal effects can drastically alter the mapping between the position of an actuator and the state of the system. 
RL \cite{suttonRL} methods are well suited to controlling systems with drift as they can learn relative and dynamic relationships rather than relying on precision actuation. 
This enhanced ability, however, comes at the cost of long training times \cite{DulacArnold2021}.
Sim-to-real methods provide an effective strategy to accelerate the learning process, where an agent is trained on a simulation and adapted to deploy in real systems\cite{Zhao2020}.
Simulated environments have been previously used to train an RL agent to align a Mach-Zehnder interferometer, reducing the number of time-consuming interactions with the experimental setup \cite{makarenko2022aligning, interferobot}. 
Work has also shown that RL-based control can help maintain the alignment of a gravitational wave detector \cite{Mukund2023}.
While sim-to-real training can save time compared to online training, it is hard to construct a simulation that captures the non-ideal behaviour of a complex system.

In the present work, we use DL-based model-free optimisation as a starting point. The data collected from this stage is then used to pre-train a model-based RL algorithm. 
We demonstrate how RL dramatically improves resilience to drift when deployed on the experiment with noisy actions \cite{lea-vicky2024} and rewards.
Although the specific task is the mode-matching of an optical resonator, the data-driven, task-agnostic nature of our method allows it to be generalised to many optical systems where suitable observables and controls are available.

\begin{figure}
    \centering
    \includegraphics[width=1.0\textwidth]{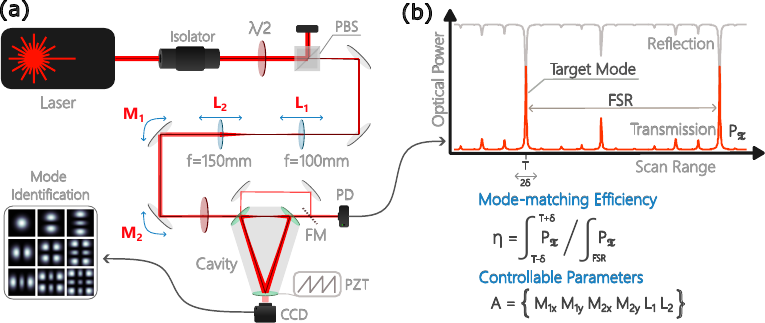}
    \caption{(a) The experimental setup provided to the ML agents, where the position of two lenses adjusts the beam waist and the angles of two mirrors steer the beam into the cavity. The cavity length is modulated using a piezo (PZT) attached to the rear mirror. The reflected beam can be cast onto the detector using a flip mirror (FM), as shown. (b) Reflected and transmitted cavity spectra, where the photodetector power shows the modes (peaks) present for different scan values. A target mode is defined by the operator, from which the mode matching efficiency is calculated using the depicted equation as an approximation to Eq.\ref{cost-func}.}
    \label{exp}
\end{figure}

\section{Experimental System}
Optical cavities, or resonators, consist of two or more highly reflective mirrors positioned to trap light through repeated reflections and refocusing of an input laser beam. 
As the input light beam travels around the cavity, they interfere constructively on satisfying the resonant condition dictated by the cavity length. This leads to the formation of various transverse electromagnetic (TEM) modes, with the intensity distribution described by the Hermite-Gaussian function \cite{siegman86}. 
By mounting one of the cavity mirrors on a piezoelectric actuator, the cavity length can be varied.

The input beam, assuming it is not perfectly aligned, can be decomposed into a linear combination of the cavity eigenmodes, which in general are not frequency degenerate. On detecting the transmitted power, one finds a spectrum of cavity modes as the length of the resonator is scanned, the composition of which is determined by the misalignment of the input beam. Thus, these spectra are commonly used as a diagnostic tool for cavity alignment.  The pattern of modes repeats every time the cavity round-trip length changes by one wavelength of the input, giving rise to the free spectral range (FSR) of the resonator as shown in Fig.~\ref{exp}(b).
Aligning a beam to a cavity requires a pair of optics (usually mirrors) to control the angular and transverse displacement of the beam. Mode-matching to a cavity requires a system of lenses to adjust the beam's waist to overlap with the principal mode and position it at a point in space defined by the cavity's geometry. 
Depending on the geometry, stability, and size of the cavity, the difficulty of this task can vary wildly, as is the case with many FSO alignment tasks.

\subsection{Design and Implementation}

The optical setup required to demonstrate automation of alignment and mode-matching to a cavity is shown in Fig.\ref{exp}. The incident light is derived from a Diablo NPRO 1064 nm laser. 
We used a triangular ring cavity with an estimated finesse of $\mathcal{F} = 130$ and $1/e^2$ beam diameter of 450~$\mu$m. It consisted of two plane mirrors and one spherical mirror, forming a travelling wave resonator.
Two lenses $(L_1, L_2)$ of focal lengths 100 mm and 150 mm were mounted on actuator-driven translation stages (Thorlabs Z825B). 
Following this, two mirrors with actuators (Thorlabs Z812) allowed adjustment of angles along both axes $(M_{1X}, M_{1Y}, M_{2X}, M_{2Y})$. This allowed the steering mirrors to translate the beam by $\sim$~6 mm, corresponding to a complete misalignment of the input beam.
The six actuators were individually driven by SparkFun motor drivers, controlled by an FPGA (NI PXI-7813R), and interfaced to LabVIEW, which also monitored their position via the actuator encoders.
A 5~Hz sawtooth signal was generated by a Moku:Go to scan the piezo, which was amplified to accommodate at most 1.5 FSRs, allowing control over the phase, amplitude, and offset of the scan signal.

A photodetector (PD) measured the intensity spectrum of the cavity transmission. 
In addition, a CCD was placed on the backside of the cavity to image the spatial modes that arise from the misalignment of the input beam and to identify the target mode.
The collection of the PD signal was triggered by the Moku:Go synced with the piezo scan. 
The cavity spectrum is the primary feedback from the experiment to the ML agents.
The acquisition and control system was unified through a Python interface, as illustrated in Fig.\ref{scheme}, which was also responsible for signal processing, peak detection, and calculating the degree of achieved alignment.

\subsection{Setting a goal} \label{sec:goal}
When coupling a beam into a resonator, the mode matching efficiency, $\eta$, is a useful metric of success. Mathematically, $\eta$ is described as the overlap integral of the complex electric field of the input beam $E_{in}$ and the target mode $E_{T}$ divided by the product of their intensities. 
Experimentally, $\eta$ can be estimated by looking at the cavity mode spectrum, like the one illustrated in Fig.~\ref{exp}(b). 
Comparing the total area under the mode spectrum to the area of the target mode, we are able to find $\eta$ using the transmitted spectra, $P_\mathfrak{T}$.
Given a mode spectrum that spans a single FSR, we may define the theoretical and experimental relations as,

\begin{equation}
    \label{cost-func}
        \eta = \frac{|\int E_{in}^*E_{T}\,dA|^2}{\int |E_{in}|^2\,dA \int |E_{T}|^2\,dA}
        = \frac{\int_{T} P_\mathfrak{T}}
        {\int_{\mathrm{FSR}} P_\mathfrak{T}}
\end{equation}

\noindent where the integration area, $A$, spans the entire beam cross-section and the limits on the integration of the experimentally measured powers are illustrated in Fig.~\ref{exp}(b). 
On transmission, the output mode has fixed alignment. If the cavity is too misaligned, however, there are no modes to observe, and $\eta$ cannot be calculated.
In a practical setting, a coarse raster scan of the system can be used to recover a usable transmitted spectrum in the event of extreme misalignment. 
The relation between $\eta$ and the transmitted mode is used as a reward function to be maximised by the ML agents trying to reach goal alignment.
In the current work, we seek to optimise the purity of the TEM$_{00}$ mode by using the cavity as a ``mode cleaner'' that passes just TEM$_{00}$ and rejects any higher-order modes. This mode is a natural choice since it is most commonly used in optical experiments, although, in principle, any mode could be optimised with our method. 

The expression for $\eta$ in terms of the measured power will only reach a maximum value of 1 under ideal conditions.  
The value of $\eta$ found by integrating the raw data from the detector suffers from a combination of fluctuating input power, integrated detector noise and offset levels. This limited the maximum value of $\eta$ to $0.6\pm0.05$ when averaged across all experimental runs. 
While a maximum in $\eta$ will correspond closely to a maximum in the physical mode matching efficiency, the uncertainty in $\eta$ renders comparison between experimental runs difficult. We are able to correct for this in post-processing using an algorithm based on peak detection, which provides a more accurate estimate of the mode matching efficiency. We term this corrected value $\eta'$ and use this for comparison across the entire experimental data. In our experiment, the best mode matching we observed for manual and ML alignment is $\sim$ $0.96$ and $0.97$, respectively.

\subsection{Drifts in the System}
\label{drifts}
A challenging addition to any system undergoing optimisation and control is drift. We observe two main sources of drift in our system. The most prominent effect is the actuators failing to reach the same physical position for a given parameter value. This is caused by accumulating backlash errors and inherent hysteresis. Due to the limitations of the actuator hardware, this cannot be corrected. Instead, the ML agents must operate with this limitation. We also observe that due to environmental temperature fluctuations, the cavity peak position for a given piezo scan voltage will drift over short time scales. This is automatically corrected in post-processing, using an autocorrelation method, and thus is not observed by the ML agents. 

\begin{figure}
    \includegraphics[width=\textwidth]{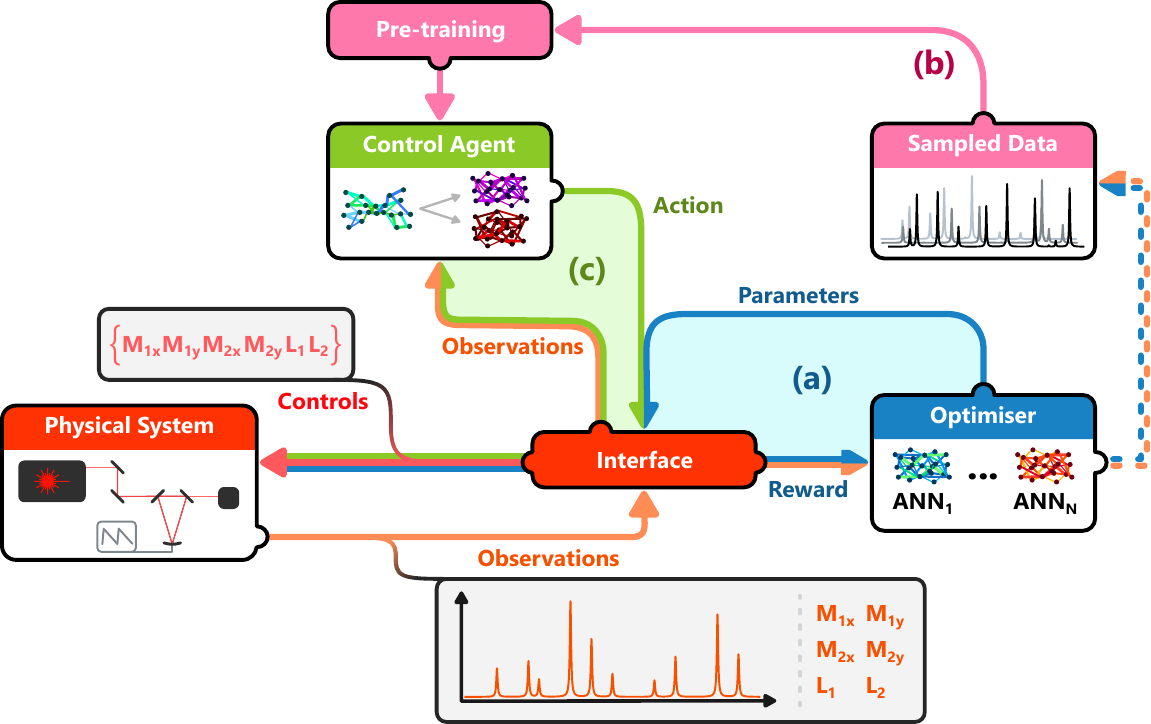}
    \caption{Schematic of the machine learning methods used. The unified Python interface serves as the common access point to the experimental system. Illustrated as three processes: (a) a parameter optimisation routine as described in Sec.\ref{sann}, identifying the optimal region; (b) data sampled during the optimisation is used to pre-train the RL control agent; (c) the control agent uses pre-trained models to act on the physical system through an interface as described in Sec.\ref{aqua}.}
    \label{scheme}
\end{figure}

\section{Machine Learning for Optimisation and Control}  
The reward function defined in Eq.~\ref{cost-func} provides the ML agents with a measure of the system's performance. In this case, maximising the reward will lead to optimised mode matching. Ideally, the relation between the experimental parameters and the reward would be static, such that the system behaviour is repeatable. 
Once trained, a model would have a fixed mapping between the actuator positions and the reward, allowing near-instant alignment.
Perfect static mapping is, however, not particularly realistic. Most experimental systems are prone to some drift, and re-evaluation of the same parameters will often not return the same reward.

A human experimentalist is never aware of the precise angle and position of actuators under their control. Rather, they make relative adjustments to the angles of the mirrors and positions of the lenses, observing changes in the mode spectrum. This approach works regardless of any drift, provided the alignment procedure is fast compared to the rate of drift. 
Building on this idea, a control agent that does not rely on a static relationship but instead makes relative adjustments (actions), by accessing the degree of misalignment through observations can mimic an experimentalist. RL methodologies are suitable for such interactive scenarios to achieve tolerance against experimental drift. 
In this paradigm, an ML agent uses an observation from the experiment to choose an action that maximises a returned reward. In our case, the transmitted cavity spectrum serves as the observation, and the corresponding value of $\eta$ serves as the reward.

In general, RL algorithms require vast amounts of training data and online interaction to learn a particular task \cite{DulacArnold2021}. For many experimental and physical systems, the data requirements of traditional RL algorithms are prohibitive. 
Recent advances in RL and generative ML have demonstrated an ability to learn complex environmental relationships from observational data using ``world models'' \cite{world-models, planet}, which can be used to develop control strategies. 
Planning algorithms that use model-based iterative search \cite{cem} to look for an optimal action have been shown to significantly improve data efficiency by maximising computational usage.
Building on these ideas, we present a staged approach to optimisation and control of the cavity system, which can extend to other physical systems. 
Firstly, a static online-DL optimisation is applied to the system, similar to our previous work on atomic cooling \cite{Tranter2018}. 
Secondly, the data from optimisation is used to train a model-based planning algorithm. Thirdly, after some initial pre-training, the planning agent is deployed on the system to implement RL-based online control. 
In the following sections, we describe each stage of the approach and their respective utility.

\subsection{Optimisation with an Ensemble of Learners}
\label{sann}
Neural Networks (NNs), backed by the universal approximation theorem \cite{Hornik1989}, are capable of approximating complex relationships between inputs and outputs. It allows them to effectively act as a proxy or surrogate model for computationally expensive simulations or complex physical systems. 
If a model is capable of providing an accurate mapping between the control parameters and a system's performance, then efficient optimisation of this model can be performed, identifying potential optimal parameters. 
These parameters are tested on the system in an online-learning setting before the surrogate model is retrained with the updated information, and the model is again queried for an optimal point. This process is depicted in Fig. \ref{scheme}(a) as a feedback loop between an ensemble of NNs and the experiment, where a new set of parameters is provided for testing and a resultant reward is returned. The stochastic artificial neural network (SANN) technique has been successfully demonstrated in cold atomic systems \cite{Tranter2018, Gupta2022}. Of particular note is the sample efficiency of this method, which rapidly optimises in excess of 60 individual control parameters.
In the present work, the SANN is tasked with performing the initial system optimisation. This is a static method that cannot account for rapid system drift. Nevertheless, it provides data that is suitable for training the RL-based control algorithm, as depicted by Fig.~\ref{scheme}(b). While raw cavity spectra are also recorded during this process, outside of reward calculation, they are not explicitly used by the SANN.

\subsection{Model-based Planning and Control}
\label{aqua}
Observations play a crucial role in RL as they provide the control agent with the context of the environment's state, enabling informed decision-making. 
It may include measurements from instruments, sensor readings, or anything experimentally accessible.
In our case, the cavity transmission spectrum provides information about how the input beams are aligned and mode-matched, partially revealing the system's internal state. 
Observing the changes in the spectrum to given actions allows the agent to intuit the correct adjustment to apply to the control parameters. Crucially, this relative change is independent of the system drift, as the observation quantifies the system's state, including drift. To this extent, the spectra and control parameters can be used to model the dynamics of the experiment.

Given the nature of tasks involving sequential decision-making, conventional RL techniques usually reward an agent over multiple actions.
We simplify this process by maximising the reward solely for the next action, assuming that nothing experimentally restricts us from reaching the goal in a single action.
To avoid the need to craft a unique feature extractor for each unique cavity or system, we leave the task to unsupervised learning with generative models. We encode the cavity spectra and their associated control parameters into a lower-dimensional feature space, referred to as the latent state \cite{vae}.
This becomes our approximation of the hidden internal state of the system and allows the agent to predict the next state and the associated reward for a given action.
States are characterised by both observations and parameters and will respond to changes caused by drift, which alters future predictions. 

Starting with an observation, the search for the optimal next action occurs in the latent space, using a planner guided by a policy \cite{POPLIN}. 
Given a state, the policy maps to the best possible action based on available data, effectively imitating \cite{Hussein2017} an expert agent that is expected to reach the goal state in a single step.
We henceforth refer to this agent formed collectively of the prediction models, policy, and the planner as AQUA (A Quantised Utility Agent).
The models used can be pre-trained using data collected by the SANN without any additional interaction with the experiment, providing them with a priori knowledge of the system.
The action chosen by the planner is applied to the experiment, and the process is repeated until the target reward or episode length is met (see Fig.~\ref{scheme}(c)). An episode is defined as a collection of steps taken by AQUA before a stopping condition is satisfied.
The target reward is set using a threshold, defining the goal and stopping criteria for an episode, which is updated to reach a higher reward upon successful completion.
When deployed online, the models are periodically retrained with newly collected data, retaining their accuracy over time.

\begin{figure}
    \includegraphics[width=\textwidth]{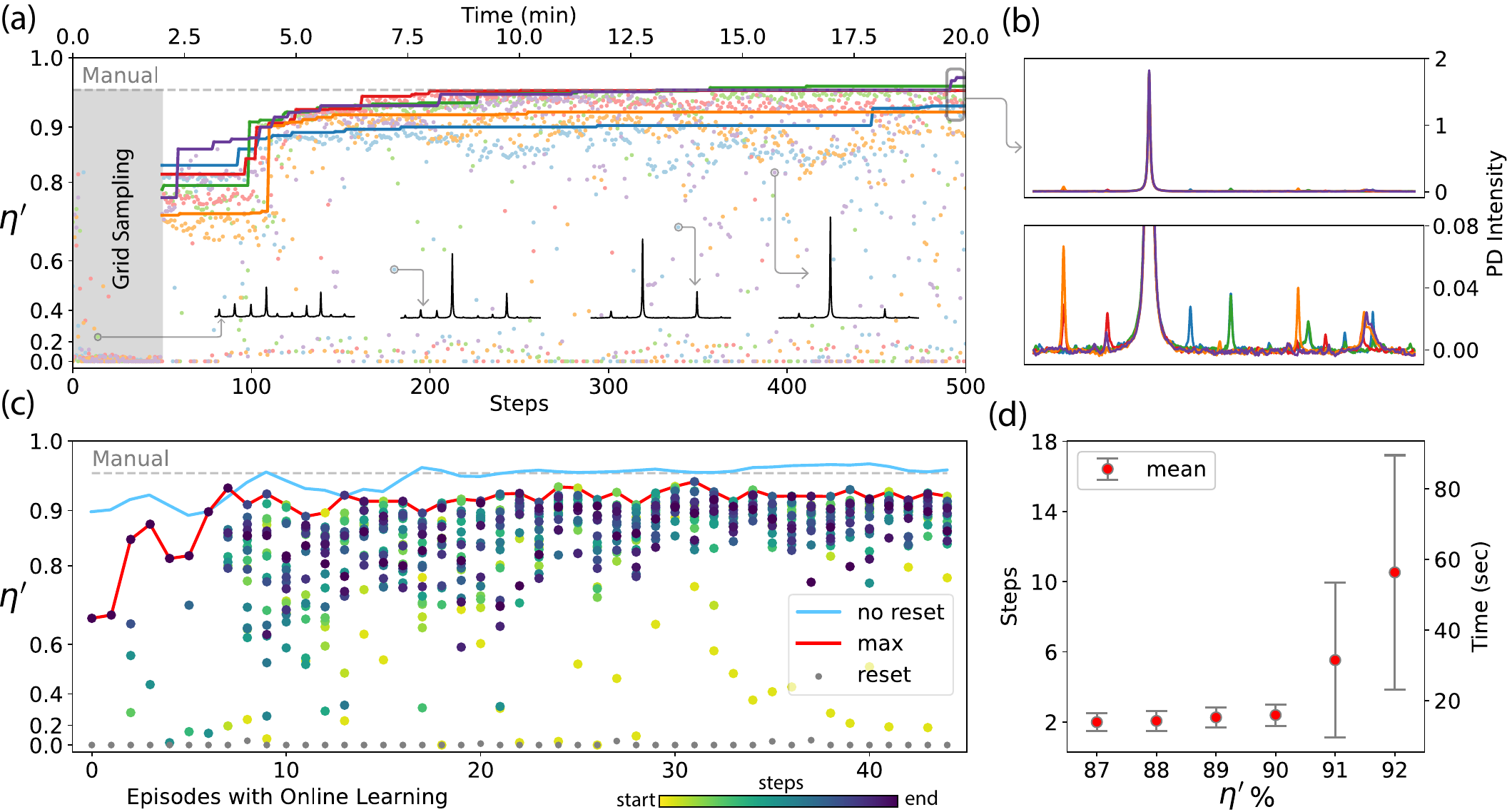}
    \caption{Performance of the machine learning methods applied to the optical cavity system. 
    (a) The performance of the SANN agent, with each colour in the scatter plot showing the individual steps for a unique experimental run. The corresponding lines indicate the highest observed mode matching till that step. The insets show the evolution of the cavity spectrum as the agent converges.
    (b) Overlay of the observed cavity spectra with the highest mode matching efficiency. The bottom plot shows a zoomed region that illustrates the remaining undesired modes.
    (c) The performance of AQUA when deployed online after pretraining. Starting from a random reset point, the agent learns to align the cavity over a few steps ($\leq20$) in an episodic fashion. Individual steps are coloured such that the episode proceeds according to the colour bar. The blue line demonstrates the maximum achieved $\eta'$ for an episode when starting from the previous episode's final position instead of performing a reset.
    (d) Distribution of steps taken to realign the system from a random reset point to a minimum required value of $\eta'$. 
    The realignment time, dictated by actuator speed, is also shown on the right vertical axis.}
    \label{results}
\end{figure}

\section{Results} \label{sec:results}
For the initial optimisation, as described in Sec.~\ref{sann}, a SANN with 6 identical NNs was used to optimise the alignment of the cavity from some arbitrary starting position. After an initial grid sampling to provide the models with some base system knowledge, the NNs were principally in charge of predicting optimal control parameters.  

Figure~\ref{results}(a) illustrates a subset of five unique experimental runs performed on different days. 
For comparison across multiple experimental runs, we use the physical mode matching efficiency $\eta'$ that compensates for the integrated detector noise, as described in section \ref{sec:goal}.
As an estimate of the maximum achievable mode matching, we performed a manual alignment before starting the optimisation runs, reaching a maximum value of $\eta'=0.96$.
The data in  Fig.~\ref{results}(a) shows that SANN could attain optimal alignment in less than 20 minutes from a position of minimal alignment, limited by the speed of actuators. In some cases, the SANN matched the maximum manual alignment value within 10 minutes.

To better see what modes remain in our spectrum, we plot the highest $\eta'$ spectra collected during each experiment, as shown in Fig.~\ref{results}(b). The remaining combinations of undesired peaks depicted in the lower plot are the result of astigmatisms in the input beam that cannot be corrected with the current experimental setup. We note that the remaining modes in each run vary, indicating that the reward function $(\eta)$ was unable to differentiate between these near-optimal alignment states.

Looking at the optimisation data, it is evident that while some runs achieved mode matching as good or better than the human effort, not all runs do so, and being in the top $5\%$ is quite rare. Furthermore, towards the end of the run, we often observe degradation in mode matching rather than improved performance. Both of these observations are driven, at least in part, by system drift, which breaks the assumption regarding a static mapping between control parameters and reward. 
When the optimiser returns to the previously recorded optimal reward region, 
a drift-induced shift in the reward space returns a lower value, contradicting previously observed information. 
There is also an issue with the consistency of our reward function, as discussed in section~\ref{sec:goal}. While easy to calculate, it is prone to fluctuations during or between experimental runs. 
In future work, we plan to experiment with alternative reward functions or filtering of the mode-matching spectra to mitigate this issue. The issues with drift and reward function noise likely also contribute to the variations in the mode spectra seen in Fig.~\ref{results}(b). We note, however, that despite the limitations of our reward function, the ML agents are still performant for the task at hand.

The data from a single SANN run was then used to pre-train AQUA as described in Sec.\ref{aqua}. This training took approximately 20 minutes.
A cavity spectrum of $\sim 1$ FSR and the associated actuator positions jointly form the observation, which was encoded into a 32-dimensional latent state. 
The pre-trained models and the planner were then deployed online in the experiment.

As previously discussed, the inherent system drift prevents an entirely repeatable response. To illustrate that the relative corrections applied as actions by AQUA are superior for control in the presence of drift, we allowed AQUA to perform constant correction on the system, with online training occurring at the end of every episode (20 steps). 
As depicted by the blue trend line in Fig \ref{results}(c), the switch to relative correction, as opposed to absolute coordinates, allows AQUA to maintain alignment approximately equal to the manual alignment maximum after $\sim 20$ episodes.  

This task, however, is relatively simple due to the misalignment state at the start of each episode being close to that of the best observed alignment. Such performance could potentially be achieved by any algorithm that can find the local optimum. To properly highlight AQUA's corrective abilities, we performed a separate experiment where the parameters were randomised (reset) at the beginning of each episode.  This data is also shown in Fig.~\ref{results}(c). 
Each reset pushed the system to a point where the observation was essentially noise ($\eta'\rightarrow 0$), mimicking an extreme case of misalignment.
For a given reset state, AQUA was then permitted a maximum of 20 steps per episode to reach a target threshold for alignment.
At the beginning of this process, the agent was given a reduced goal to reach: 70$\%$ of the highest value achieved in the preceding optimisation run. 
After a goal was reached, it was increased by 5\% until it saturated to a point where the agent could no longer exceed the goal.
Models were retrained between episodes, but only once $\geq 20$ steps had been taken. The performance until episode 7 is, therefore, purely based on the pre-trained models, reaching the goal conditions in 2 steps on average. 
On pushing the threshold beyond 100\%, we observed the agent spending more steps seeking the goal, reaching a maximum, and exploring until the episode was terminated.
We also observed a gradual reduction in the variance of rewards received during the episodes and convergent behaviour from episode 30 onwards. 
The entire experimental run duration was $\sim$ 2-3 hours with a NVIDIA Tesla P100 GPU. 
This is a substantial improvement over prior works that often take multiple days to train and converge.
More than half of this time was spent performing experimental execution being limited by actuator speeds, while the other half was spent retraining the models. As is the case with the SANN experiment, actuator backlash is stochastically accumulated during the episode steps, causing the parameter space to drift.

The data in Fig.~\ref{results}(c) shows that with resetting to a misaligned position and restricting the allowed number of steps, a maximum mode matching of $\eta'=0.94$ was achieved, which was a fairly typical performance. Fig.~\ref{results}(d) illustrates the distribution of the number of steps taken and the corresponding amount of time taken by AQUA to reach different threshold values of $\eta'$ that are comparable to human-level performance. On convergence, the agent can consistently reach the 0.92 $\eta'$ in 10 steps (< 60 seconds) on average.

Correcting for an arbitrary misalignment far from the best known configuration is challenging and often resembles a global optimisation task. AQUA can efficiently correct for these extreme cases of misalignment in a small number of steps, even in the presence of drift.
Once trained online, re-alignment can be performed by AQUA by monitoring system performance and instigating the corrective process when necessary. Model training can also continue as a background process while the system is in use. Although we illustrate the process where the agent adapts and converges when deployed to act on the system, it is not necessary to do so unless the system has drifted significantly since the last optimisation. This allows the deployment of the same pre-trained model across multiple setups.

\section{Conclusion}
Our work investigated the efficacy of online ML methods for optical alignment. The task of mode matching an optical resonator was a considered choice since these systems are known to be challenging to align, even for skilled human operators. 
By using the mode matching efficiency as a reward function, we showed that SANN was able to match the best human alignment efforts in around 10 minutes. 
This method is, however, susceptible to system drift, meaning that training data can become stale and frequent re-training would be required. 
Using a reinforcement learning system (AQUA), it was shown that near-optimal alignment could be achieved in less than a minute, even when the system has drifted.

Our experiment also highlights the resilience of our ML systems, which were able to overcome noisy actions and rewards yet deliver human-like capability. Of particular note is their sample efficiency, which makes them fit for experimental systems where measurement times may be extensive. Furthermore, there is no part of AQUA that depends inherently on the particulars of an optical resonator. The approach is driven entirely by data derived from suitable observables and an informative reward function.  We plan to explore the efficacy of our system on a range of optical and quantum technologies in the near future.

\section*{Funding}
Australian Research Council (ARC) Center of Excellence for Quantum Computation and Communication Technology (CE170100012); Japan Science and Technology (JST) Agency (Moonshot R$\&$D, Grant No. JPMJMS2064).

\section*{Acknowledgements}
A.S., V.V., T.M., S.A, P.K.L., B.B. and A.T. acknowledge the support of the ARC. B.C. and A.F. acknowledge the support of the JST.

\section*{Disclosures}
The authors declare no conflicts of interest.

\section*{Data availability}
Data underlying the results presented in this paper are not publicly available at this time but may be obtained from the authors upon reasonable request.

\bibliography{sample}

\end{document}